\documentclass[letter,10pt,conference]{IEEEtran}
\usepackage{pifont}
\usepackage{amsfonts}
\usepackage{cite}%
 \usepackage{enumitem}

 \usepackage{relsize}
\usepackage{cite}
\usepackage{amsfonts}
\usepackage{mathrsfs}
\usepackage{bbm}
\usepackage{pifont}
\usepackage{amsfonts}
\usepackage{amsmath, amsthm}
\usepackage{tabularx}
\usepackage{listings}
\usepackage{graphicx}
\usepackage{float}
\usepackage{amssymb}
\usepackage{array}
\usepackage{fancyhdr}
\usepackage{cases}
 \usepackage{supertabular}
\usepackage{url}
\usepackage{fancyhdr}

\usepackage{psfrag}
\usepackage{color}
\usepackage{cite}
\usepackage{bbding}
\newcommand{\bcap} {\hspace{2pt} \mathlarger{\cap}
\hspace{2pt}}
\newcommand{\qeda}{\hfill\ensuremath{\blacksquare}}

\newcommand{\bcup} {\hspace{2pt} \mathlarger{\cup}
\hspace{2pt}}

\newcommand {\C} {{\rm I\kern-5.5pt C}}

\newcommand{\bP}[1]{{\mathbb{P}}\left[{#1}\right]}
\newcommand{\bE}[1]{{\mathbb{E}}\left[{#1}\right]}

\newcommand{\1}[1]{{\bf 1}\left[#1\right]}       

\newcommand{\fsquare}{\vrule height6pt width7pt depth1pt}   
\newcommand{\myproof}{{\hfill \\ \bf Proof. \ }}           
\newcommand{\myendpf}{\hfill\fsquare \\[0.1in]}             

\def\centerhack#1{\hbox to 0pt{\hss\footnotesize #1\hss}}
\def\centerhackn#1{\hbox to 0pt{\hss #1\hss}}
\def\dchack#1{\vbox to 0pt{\vss{\hbox to 0pt{\hss#1\hss}}\vss}}

\IEEEoverridecommandlockouts

\newtheorem{lem}{Lemma}
\newtheorem{thm}{Theorem}

\newtheorem{rem}{Remark}
\newtheorem{cor}{Corollary}
\newtheorem{proposition}{Proposition}
\newtheorem*{proposition1.1}{Proposition 1.1}
\newtheorem*{proposition1.2}{Proposition 1.2}
\newtheorem*{proposition1.3}{Proposition 1.3}
\newtheorem*{proposition2.1}{Proposition 2.1}
\newtheorem*{proposition2.2}{Proposition 2.2}
\begin{document}




\title{Secure connectivity of  wireless sensor networks\\
under key predistribution with on/off channels}

\author{\IEEEauthorblockN{Jun Zhao}
\IEEEauthorblockA{{\tt junzhao@alumni.cmu.edu}\thanks{Jun~Zhao obtained his PhD from Carnegie Mellon University, Pittsburgh, PA, USA. This research was conducted when he was a postdoc with Arizona State University, Tempe, AZ, USA. \newline \indent This research was supported in part by Arizona State University, by the U.S. National Science Foundation (NSF) under Grants SaTC-1618768 and CNS-1422277, by Army Research Office under Grant W911NF-16-1-0448, and by the U.S. Defense Threat
Reduction Agency under Grant HDTRA1-13-1-0029. This research was also supported in part by CyLab and Department of Electrical \& Computer Engineering
at Carnegie Mellon University.}}}




%

\fancyhf{} 
\lhead{\small \color{red} This paper has been published in \emph{\bf International Conference on Distributed Computing Systems (ICDCS)} 2017. }
\fancyfoot[C]{\thepage}

\maketitle
\thispagestyle{fancy}\pagestyle{fancy}


%




\begin{abstract}

Security is an important issue in wireless sensor networks (WSNs), which are often deployed in hostile environments. The $q$-composite key predistribution scheme has been recognized as a suitable approach to secure WSNs. Although
the $q$-composite scheme has received much attention in the literature, there is still a lack of rigorous analysis for secure
WSNs operating under the $q$-composite scheme in consideration of the unreliability of   links.  One main difficulty lies in analyzing the network topology whose links are not independent. Wireless links can be unreliable in practice due
to the presence of physical barriers between sensors or because
of harsh environmental conditions severely impairing communications. In this paper, we resolve the difficult challenge and
investigate $k$-connectivity in secure WSNs
operating under the $q$-composite scheme with unreliable
communication links modeled as independent on/off channels, where $k$-connectivity ensures connectivity despite the failure of any $(k-1)$ sensors or links, and connectivity means that any two sensors can find a path in between for secure communication.  Specifically, we derive the asymptotically exact probability and a zero-one law for $k$-connectivity. We further use the theoretical results to provide   design guidelines for secure WSNs.  Experimental results also confirm the validity of our analytical findings.

%
%

\end{abstract}

%

%


\section{Introduction}

Since Eschenauer and Gligor \cite{virgil} introduced
the basic key predistribution scheme to secure
communication in wireless sensor networks (WSNs), key predistribution schemes
have been studied
extensively in the literature over the last decade \cite{4550354,4660127,Tague:2007:CSA:1281492.1281494,4146913,Hwang:2004:RRK:1029102.1029111,6584937}. 
 The idea of  key predistribution is that cryptographic keys are assigned before deployment to ensure secure
sensor-to-sensor communications.

Among many key predistribution schemes,
the $q$-composite  scheme proposed by Chan
\emph{et al.} \cite{adrian} as an extension of the basic
Eschenauer--Gligor scheme \cite{virgil} (the $q$-composite scheme in
the case of $q=1$) has received much interest  \cite{bloznelis2013,Hwang:2004:RRK:1029102.1029111,yagan_onoff,zhao2014topological,ANALCO}.
 The $q$-composite key predistribution scheme works as follows. For a WSN with $n$
sensors, prior to deployment, each sensor is independently assigned
$K_n$ different keys which are selected \emph{uniformly at random}
from a pool $\mathcal {P}_n$ of $P_n$ distinct keys.  After deployment, any two sensors
establish a \emph{secure} link in between  \emph{if
and only if} they have at least $q$ key(s) in common \emph{and} the physical
link constraint between them is satisfied. Both $P_n$ and $K_n$ are both
functions of $n$ for generality, with the natural condition $1 \leq q <
K_n < P_n$. Examples of physical link constraints include the
reliability of the transmission channel and the distance between two
sensors close enough for communication. The $q$-composite scheme
with $q\geq 2$ outperforms the basic Eschenauer--Gligor scheme with
$q=1$ in terms of the strength against small-scale network capture
attacks while trading off increased vulnerability in the face of
large-scale attacks \cite{adrian}.

 In this
paper, we investigate $k$-connectivity in secure
WSNs employing the $q$-composite key predistribution scheme with
general $q$ under the \emph{on}/\emph{off} channel model as the
physical link constraint comprising independent channels which are
either \emph{on} or \emph{off}. A network is $k$-connected if it remains connected despite the failure of at most $(k-1)$ nodes, where nodes can fail due to adversarial attacks, battery depletion, or harsh environmental conditions \cite{ZhaoYaganGligor}; connectivity ensures that any two nodes can find a path in between \cite{yagan_onoff}.
Our results on secure $k$-connectivity include the asymptotically exact probability and also a zero--one law. The zero--one law means that the
network is securely $k$-connected with high probability under certain parameter conditions
and is not securely $k$-connected  with high probability under other parameter conditions,
where an event happens ``with high probability'' if its probability converges to $1$ asymptotically.
The zero--one law specifies the \emph{critical} scaling of the model parameters in terms of secure $k$-connectivity, while the asymptotically exact probability result provides a {\em precise} guideline for
ensuring   secure $k$-connectivity. Obtaining such a precise
guideline is particularly crucial in a WSN setting as explained below.
 To increase the chance of ($k$-)connectivity,
it is often required to increase the number of keys kept in each
sensor's memory. However, since sensors  have
 limited memory, it is desirable for practical key
distribution schemes to have low memory requirements
\cite{virgil,yagan}. Therefore, it is important to obtain the asymptotically exact probability as well as the zero--one law to
dimension the $q$-composite scheme.

%
%

Our approach to the analysis is to explore the induced random graph
models of the WSNs. As will be clear in Section~\ref{sec:SystemModel}, the graph modeling a studied WSN is an intersection
of two distinct types of random graphs. It is the intertwining
\cite{yagan_onoff,ZhaoYaganGligor}
of these two graphs that makes our analysis challenging.

We organize the rest of the paper as follows. Section
\ref{sec:SystemModel} describes the system model.
Afterwards, we detail the analytical results in Section
\ref{sec:res}. We provide   experiments in
Section \ref{sec:expe} to confirm our analytical results. Sections \ref{sec:ProofTheoremNodeIsolation} through \ref{sec-prf-upper-bounds-details} are devoted to proving the results. 
 Section \ref{related} surveys related work. Finally,  we conclude the paper in \mbox{Section \ref{sec:Conclusion}.}

 \section{System Model}
\label{sec:SystemModel}

We elaborate the graph modeling of a WSN with $n$ sensors, which
employs the $q$-composite key predistribution scheme and works under
the {on/off} channel model. We use a node set $\mathcal {V}_n = \{v_1,
v_2, \ldots, v_n \}$ to represent the $n$ sensors (the terms sensor and node are interchangeable
in this paper). For each
$v_i \in \mathcal {V}_n$, the set of its $K_n$ different keys is
denoted by $S_i$, which is uniformly distributed among all
$K_n$-size subsets of a key pool $\mathcal {P}_n$ of $P_n$ keys.

The $q$-composite key predistribution scheme is
 modeled by a uniform $q$-intersection graph  \cite{bloznelis2013,ANALCO}
denoted by $G_q(n, K_n,P_n)$, which is defined on the node set
$\mathcal{V}_n$ such that any two distinct nodes $v_i$ and $v_j$
sharing at least $q$ key(s) (an event denoted by $\Gamma_{ij}$) have
an edge in between. Clearly, event $\Gamma_{ij}$ is given by  $\big[ |S_i
\bcap S_j | \geq q \big]$, with $|A|$ denoting the
cardinality of a set $A$.

Under the {on/off} channel model, each node-to-node channel is
independently {\em on} with probability ${p_n} $ and {\em off} with
probability $(1-{p_n})$, where ${p_n}$ is a function of $n$ with
$0<{p_n}\leq 1$. Denoting by ${C}_{i j}$ the event that the channel
between distinct nodes $v_i$ and $v_j$ is {\em on}, we have
$\bP{C_{ij}} = {p_n}$, where $\mathbb{P}[\mathcal {E}]$ denotes the
probability that event $\mathcal {E}$ happens, throughout the paper.
The {on/off} channel model is represented by an Erd\H{o}s-R\'enyi
graph $G(n, {p_n})$ \cite{citeulike:4012374} defined on the node set
$\mathcal{V}_n$ such that $v_i$ and $v_j$ have an edge in between if
event $C_{ij}$ occurs.

Finally, we denote by $\mathbb{G}_{n,q}(n, K_n, P_n,
{p_n})$ the underlying graph of the $n$-node WSN operating under the
$q$-composite scheme and the on/off channel
model.  Graph
$\mathbb{G}_{n,q}(n, K_n, P_n,
{p_n})$ is defined on the node set
$\mathcal{V}_n$ such that there exists an edge between nodes $v_i$ and
$v_j$ if and only if events $\Gamma_{ij}$ and $C_{ij}$ happen at the
same time. We set event $E_{ij} : = \Gamma_{ij} \cap C_{ij}$.
Then the edge set of $\mathbb{G}_{n,q}(n, K_n, P_n,
{p_n})$ is the intersection of the edge sets of $G_q(n, K_n, P_n) $ and $G(n, {p_n})$,
 so $\mathbb{G}_{n,q}(n, K_n, P_n,
{p_n})$ can be seen as the intersection
of $G_q(n, K_n, P_n)$ and $G(n, {p_n})$, i.e.,
\begin{equation}
\mathbb{G}_{n,q}(n, K_n, P_n,
{p_n}) = G_q(n, K_n, P_n) \cap G(n, {p_n}),
 \label{eq:G_on_is_RKG_cap_ER_oy}
\end{equation}


Throughout the
paper, $q$ is an arbitrary positive integer and does not scale with
$n$.
We define $s(K_n, P_n,q)$ as the probability that two different nodes
share at least $q$ key(s) and $t (K_n, P_n,q, {p_n})$ as the probability that two
distinct nodes have a secure link in $\mathbb{G}_{n,q}(n,\hspace{-.5pt}
K_n,\hspace{-.5pt} P_n,\hspace{-.5pt} {p_n})$.   We often write $s(K_n, \hspace{-.5pt}P_n,\hspace{-.5pt}q)$ and $t (K_n,\hspace{-.5pt} P_n,\hspace{-.5pt}q,\hspace{-.5pt} {p_n})$
as $s_{n,q}$ and $t_{n,q}$ respectively for simplicity.
 Clearly,
$s_{n,q} $ and $t_{n,q}$ are the edge probabilities in graphs
$G_q(n, K_n, P_n)$ and $\mathbb{G}_{n,q}(n,
K_n, P_n, {p_n})$, respectively.  From $E_{ij} = \Gamma_{ij} \cap C_{ij}$ and the
independence of ${C}_{i j} $ and $ \Gamma_{i j} $, we obtain
\begin{align}
{t_{n,q}}  & =  \mathbb{P} [E_{i j} ]  =  \mathbb{P} [{C}_{i j} ]
\cdot \mathbb{P} [\Gamma_{i j} ] =  {p_n}\cdot
s_{n,q}. \label{eq_pre}
\end{align}

 By definition, $s_{n,q}$ is determined through
\begin{align}
s_{n,q} & =  \mathbb{P} [\Gamma_{i j} ] = \sum_{u=q}^{K_n}
  \mathbb{P}[|S_{i} \cap S_{j}| = u] , \label{psq1}
\end{align}
where we derive $\mathbb{P}[|S_{i} \cap S_{j}| = u]$ as follows.

Note that $S_{i}$ and $S_{j}$ are independently and uniformly
selected from all $K_n$-size subsets of a key pool with size $P_n$.
Under $(|S_{i} \bcap S_{j}| = u)$, after $S_i$ is determined, $S_j$
is constructed by selecting $u$ keys out of $S_i$ and $(K_n-u)$ keys
out of the key pool $\mathcal {P}_n$. Hence, if $P_n \geq 2K_n$ and $K_n \geq q$, we have
\begin{align}
\mathbb{P}[|S_{i} \cap S_{j}| = u] & =
\frac{\binom{K_n}{u}\binom{P_n-K_n}{K_n-u}}{\binom{P_n}{K_n}},
\quad\text{for } u =1,2,\ldots, K_n,
\label{u3}
\end{align}
which along with (\ref{eq_pre}) and (\ref{psq1}) yields
\begin{align}
t_{n,q} = {p_n}\cdot  \sum_{u=q}^{K_n}
\frac{\binom{K_n}{u}\binom{P_n-K_n}{K_n-u}}{\binom{P_n}{K_n}}. \label{u4}
\end{align}

Asymptotic expressions of $s_{n,q}$ and $t_{n,q}$ can also be given. If
$\frac{{K_n}^2}{P_n} = o(1)$, we obtain from Lemma \ref{lem_eval_psq} or \cite[Lemma 1]{zhao2014topological} that $s_{n,q}
 \sim \frac{1}{q!}\big(\frac{{K_n}^2}{P_n}\big)^q $, which with (\ref{eq_pre}) leads to $$t_{n,q} \sim {p_n}\cdot
 \frac{1}{q!}\bigg(\frac{{K_n}^2}{P_n}\bigg)^q.$$ In the above results, for two positive sequences $f_n$ and
$g_n$, the relation $f_n \sim g_n$ means $\lim\limits_{n \to
  \infty}({f_n}/{g_n})=1$; i.e., $f_n$
  and $g_n$ are asymptotically equivalent.

\section{The Results} \label{sec:res}

We present and discuss our results in this section.
The
natural logarithm function is given by $\ln$. All limits are understood with $n
\to
  \infty$.  
 We use the standard
asymptotic notation $o(\cdot), O(\cdot), \Omega(\cdot), \omega(\cdot),
\Theta(\cdot), \sim$; see \cite[Page 2-Footnote 1]{ZhaoYaganGligor}.

Theorem \ref{thm:OneLaw+NodeIsolation} below presents the asymptotically exact probability and a zero--one  law for connectivity in a  graph $\mathbb{G}_{n,q}(n, K_n,P_n, {p_n})$.

\begin{thm}  \label{thm:OneLaw+NodeIsolation}

For a graph $\mathbb{G}_{n,q}(n, K_n,P_n, {p_n})$, with a \mbox{sequence}  $\alpha_n$ defined through
\begin{align}
t_{n,q}  & = \frac{\ln  n  + (k-1) \ln \ln n  +
 {\alpha_n}}{n},   \label{eq:scalinglaw}
\end{align}
where $t_{n,q}$ is given by (\ref{u4}), then it holds under $ K_n =
\Omega(n^{\epsilon})$ for a positive constant $\epsilon$, $ \frac{{K_n}^2}{P_n}  =
 o\left( \frac{1}{\ln n} \right)$,  and $ \frac{K_n}{P_n} = o\left( \frac{1}{n\ln n} \right)$ that
\begin{align}
& \hspace{-35pt}  \lim_{n \rightarrow \infty }  \mathbb{P}\big[\mathbb{G}_{n,q}(n, K_n,P_n, {p_n}) \mbox{ is $k$-connected.}\big]
 \nonumber
\\ &  \hspace{-35pt}  =  e^{- \frac{e^{-\lim_{n \to \infty} \alpha_{_n}}}{(k-1)!}} \label{thm-mnd-alpha-finite-kcon-compact}
\end{align}
\vspace{-5pt}
\begin{subnumcases}{ \hspace{-10pt} =}
e^{- \frac{e^{-\alpha ^*}}{(k-1)!}}, &\textrm{if } $\lim\limits_{n \to \infty} \alpha_n = \alpha ^* \in (-\infty, \infty)$,\label{thm-mnd-alpha-finite-kcon} \\
1, & \textrm{if } $\lim\limits_{n \to \infty} \alpha_n =  \infty$, \label{thm-con-eq-1}\\
0, & \textrm{if } $\lim\limits_{n \to \infty} \alpha_n = - \infty$.  \label{thm-con-eq-0}
\end{subnumcases}
\end{thm}

%

%
%

For $k$-connectivity in
$\mathbb{G}_{n,q}(n, K_n,P_n, {p_n})$, the result (\ref{thm-mnd-alpha-finite-kcon-compact}) of Theorem \ref{thm:OneLaw+NodeIsolation} presents the asymptotically exact
probability, while (\ref{thm-con-eq-0}) and (\ref{thm-con-eq-1}) of Theorem \ref{thm:OneLaw+NodeIsolation} together constitute a zero--one law, where a zero--one law means that the probability of a graph having a certain property asymptotically converges to $0$ under some conditions and to $1$ under some other conditions.  The result (\ref{thm-mnd-alpha-finite-kcon-compact}) compactly summarizes (\ref{thm-mnd-alpha-finite-kcon})--(\ref{thm-con-eq-0}).

Theorem \ref{thm:OneLaw+NodeIsolation} shows that the critical scaling of $t_{n,q}$ for $k$-connectivity in graph $\mathbb{G}_{n,q}(n, K_n,P_n, {p_n})$ is $\frac{\ln  n + (k-1) \ln \ln n}{n}$. The conditions in Theorem \ref{thm:OneLaw+NodeIsolation} are enforced
merely for technical reasons, but they are practical and often hold  in realistic
wireless sensor network applications \cite{adrian,DiPietroMeiManciniPanconesiRadhakrishnan2004,virgil}. More specifically, the condition on $K_n$ (i.e., $ K_n =
\Omega(n^{\epsilon})$) is less appealing but is not much a problem because $\epsilon$ can be arbitrarily small. In addition,  $\frac{{K_n}^2}{P_n}=
 o\left( \frac{1}{\ln n} \right) $ and $ \frac{K_n}{P_n} = o\left( \frac{1}{n\ln n} \right)$
hold in practice since the key pool size $P_n$ grows at least linearly with $n$ and is expected to
be several orders of magnitude larger than the key ring size $K_n$ (see \cite[Section 2.1]{virgil} and \cite[Section III-B]{yagan_onoff}).

%
%

 Below, we first provide experimental results before proving Theorem \ref{thm:OneLaw+NodeIsolation} in detail.
\begin{figure}[!t]
\hspace{-30pt} \includegraphics[width=0.56\textwidth]{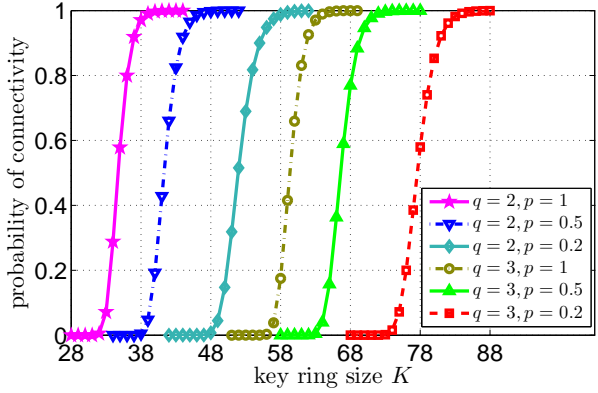}  \vspace{-8pt} \caption{Empirical probability
that $\mathbb{G}_{n,q}(n, K, P,  {p})$
         is connected as a function of $K$ for $q=2,3$ and
         $p=0.2, 0.5, 1$
         with $n=1,000$ and $P=10,000$. In each case, the empirical probability value is
obtained by averaging over $500$ experiments.
}
 \label{figure:connect}
\end{figure}

\section{Experimental Results}
\label{sec:expe}

We now present experiments to confirm Theorem \ref{thm:OneLaw+NodeIsolation}.

In the experiments, we fix the number of nodes at $n=1,000$ and the key pool size at $P=10,000$. We specify the required amount $q$ of key overlap as $q= 2, 3$, and the probability $p$ of an channel being \emph{on} as $p=0.2, 0.5, 1$, while varying the parameter $K$ from $28$ to $88$.
For each parameter pair $(q, p, K)$, we generate $500$
independent samples of the graph $\mathbb{G}_{n,q}(n, K,P, {p})$ and count the number of times (out of a
possible $500$) that the obtained graphs are connected. Dividing the counts by $500$, we obtain the
empirical probabilities for connectivity.

In Figure \ref{figure:connect},
 we depict the resulting empirical
probability of connectivity in $\mathbb{G}_{n,q}(n, K, P,  {p})$ versus $K$.
 From Figure \ref{figure:connect}, the threshold behavior of the probability of
connectivity is evident from the plots. Based on (\ref{u4}) and (\ref{eq:scalinglaw}), we also compute the minimum integer value of
$K^*$ that satisfies
\begin{equation}
t (K^*, P,q, {p}) = {p}\cdot  \sum_{u=q}^{K^*}
\frac{\binom{K^*}{u}\binom{P-K^*}{K^*-u}}{\binom{P}{K^*}} >
 \frac{\ln n}{n}. \label{eq:threshold}
\end{equation}
For the six curves in Figure 1, from leftmost to rightmost, the corresponding $K^*$ values
are 35, 41, 52, 60, 67 and 78, respectively. Hence, we see that the connectivity
threshold prescribed by (\ref{eq:threshold}) is in
agreement with the experimentally observed curves for
connectivity.

\section{Basic Ideas for Proving Theorem
\ref{thm:OneLaw+NodeIsolation}}
\label{sec:ProofTheoremNodeIsolation}



The basic ideas to show Theorem \ref{thm:OneLaw+NodeIsolation} are as follows. We decompose the theorem results into lower and upper bounds, where the lower bound  is proved by associating our studied  graph intersection $\mathbb{G}_{n,q}$ (i.e., $G_q(n, K_n, P_n) \cap G(n, {p_n})$) with an Erd\H{o}s--R\'enyi graph, while the upper bound  is obtained by associating the studied $k$-connectivity property in Theorem \ref{thm:OneLaw+NodeIsolation} with minimum node degree.

\subsection{{Decomposing the results into lower and upper bounds}} \label{sec-Decomposing}

Note that in Theorem \ref{thm:OneLaw+NodeIsolation}, the results (\ref{thm-mnd-alpha-finite-kcon})--(\ref{thm-con-eq-0}) are compactly summarized as (\ref{thm-mnd-alpha-finite-kcon-compact}); i.e., $\lim\limits_{n \to \infty}\bP{\text{$\mathbb{G}_{n,q}\iffalse_{on}^{(q)}\fi$  is $k$-connected.}}=e^{- \frac{e^{-\lim_{n \to \infty} \alpha_n }}{(k-1)!}}$. To prove (\ref{thm-mnd-alpha-finite-kcon-compact}) via decomposition, we show that the probability $\mathbb{P}[\hspace{2pt}\mathbb{G}_{n,q}\text{ is
  $k$-connected.}\hspace{2pt}]$ has a lower bound $e^{-
\frac{e^{- \iffalse \lim\limits_ \fi \lim_{n \to \infty}{\alpha_n}}}{(k-1)!}} \times [1 -o(1)]$ and an upper bound $e^{-
\frac{e^{- \iffalse \lim\limits_ \fi \lim_{n \to \infty}{\alpha_n}}}{(k-1)!}} \times [1 + o(1)]$, where a sequence $x_n$ can be written as $o(1)$ if $\lim_{n \to \infty} x_n = 0$. Afterwards, the obtained (\ref{thm-mnd-alpha-finite-kcon-compact}) implies (\ref{thm-mnd-alpha-finite-kcon})--(\ref{thm-con-eq-0}).

\subsection{{Proving the lower bound by showing that our graph intersection $\mathbb{G}_{n,q}$ contains an Erd\H{o}s--R\'enyi graph}} \label{sec-prf-lower-bounds}

To prove the lower bound of $k$-connectivity in our studied graph intersection $\mathbb{G}_{n,q}$ (i.e., $G_q(n, K_n, P_n) \cap G(n, {p_n})$), we will show that the studied graph $\mathbb{G}_{n,q}$ contains an Erd\H{o}s--R\'enyi graph as its spanning subgraph with probability $1-o(1)$, and   show that the lower bound also holds for the Erd\H{o}s--R\'enyi graph. More specifically, the Erd\H{o}s--R\'enyi graph under the corresponding conditions is $k$-connected with probability \mbox{$e^{-
\frac{e^{- \iffalse \lim\limits_ \fi \lim_{n \to \infty}{\alpha_n}}}{(k-1)!}} \times [1-o(1)]$} .

We give more details for the above idea in Section \ref{sec-prf-lower-bounds-details}.

\subsection{{Proving the upper bound by considering minimum node degree}} \label{sec-prf-upper-bounds}

To prove the upper bound of $k$-connectivity in our studied graph $\mathbb{G}_{n,q}$, we leverage the necessary condition on the minimum (node) degree enforced by $k$-connectivity, and explain that the upper bound also holds for the requirement of the minimum degree. Specifically, because a necessary condition for a graph to be $k$-connected is that the minimum degree is at least $k$ \cite{Citeulike:505396}, $\bP{\text{$\mathbb{G}_{n,q}\iffalse_{on}^{(q)}\fi$  has a minimum degree at least $k$.}}$ provides an upper bound for $\bP{\text{$\mathbb{G}_{n,q}\iffalse_{on}^{(q)}\fi$  is $k$-connected.}}$. We will prove that $\bP{\text{$\mathbb{G}_{n,q}\iffalse_{on}^{(q)}\fi$  has a minimum degree at least $k$.}}$ is upper bounded by $e^{-
\frac{e^{- \iffalse \lim\limits_ \fi \lim_{n \to \infty}{\alpha_n}}}{(k-1)!}} \times [1 + o(1)]$ so it becomes immediately clear  that $\bP{\text{$\mathbb{G}_{n,q}\iffalse_{on}^{(q)}\fi$  is $k$-connected.}}$ is also upper bounded by $e^{-
\frac{e^{- \iffalse \lim\limits_ \fi \lim_{n \to \infty}{\alpha_n}}}{(k-1)!}} \times [1 + o(1)]$.

We give more details for the above idea in Section \ref{sec-prf-upper-bounds-details}.

In addition to the arguments above, we also find it useful to confine the deviation  $\alpha_n$ in Theorem  \ref{thm:OneLaw+NodeIsolation}. We discuss this idea as follows.

\subsection{Confining the deviation  $\alpha_n$ in Theorem  \ref{thm:OneLaw+NodeIsolation}} \label{sec-Confining}

We will show that to prove Theorem \ref{thm:OneLaw+NodeIsolation}, the deviation $\alpha_n$ in the theorem statement can  be confined as $\pm  o(\ln n)$. More specifically, if Theorem \ref{thm:OneLaw+NodeIsolation} holds under the extra condition $|\alpha_n |= o(\ln n)$, then Theorem \ref{thm:OneLaw+NodeIsolation} also holds regardless of the extra condition. This extra condition will be useful for the aforementioned steps in Sections \ref{sec-prf-lower-bounds} and \ref{sec-prf-upper-bounds}. We present more details for the above idea in the next section.

\section{Confining the Deviation $|\alpha_n|$ as $o(\ln n)$ in Theorem \ref{thm:OneLaw+NodeIsolation}}

In this section, we show that
the extra condition $|{\alpha_n} |=  o ( \ln n)$
can be introduced in proving
Theorem \ref{thm:OneLaw+NodeIsolation}, where $|{\alpha_n} |$ is the absolute value of $\alpha_n$. Since ${\alpha_n}$ measures the deviation of the edge probability $t_{n,q}$
from the critical scaling $\frac{ \ln n + (k-1)\ln \ln n }{n}$, we call the extra condition $|{\alpha_n} |=  o ( \ln n)$ as \emph{the confined deviation}.
Then our goal here is to show\vspace{-2pt}
\begin{align}
\text{Theorem \ref{thm:OneLaw+NodeIsolation} with the confined deviation} ~~\Rightarrow ~~
\text{Theorem \ref{thm:OneLaw+NodeIsolation}}.
\label{with_extra}
\end{align}

We write $\mathbb{G}_{n,q}$ back as $G_q(n, K_n, P_n) \bcap G(n, p_n)$ based on (\ref{eq:G_on_is_RKG_cap_ER_oy}),  and  write $t_{n,q}$ (i.e., ${t (K_n, P_n,q, {p_n})} $) back as $s(K_n, P_n,q) \times p_n $ based on (\ref{eq_pre}).

  \begin{lem} \label{lem_Gq_cplinga}
 {
For a graph $G_q(n, K_n, P_n) \bcap G(n, p_n)$   on a probability space $\mathbb{S}$ under
\begin{align}
& \begin{array}{l} \textstyle{\frac{{K_n}^2}{P_n} = o\left( \frac{1}{\ln n} \right),  \frac{K_n}{P_n} = o\left( \frac{1}{n\ln n} \right) \text{and }} \\ \textstyle{K_n = \Omega(n^{\epsilon})\text{ for a positive constant }\epsilon} \end{array} \label{conditions-lem_Gq_cplinga} \\
&\text{\,~(i.e., the conditions of Theorem \ref{thm:OneLaw+NodeIsolation}),} \nonumber
\end{align}
 with a sequence $\alpha_n$ defined by $s(K_n, P_n,q) \times p_n  = \frac{\ln  n + {(k-1)} \ln \ln n + {\alpha_n}}{n}$, the following results hold:
\begin{itemize}[leftmargin=12pt]
\item[(i)] If ${\lim_{n \to \infty}\alpha_n = \infty}$, there exists   a graph $G_q(n, K_n, P_n) \bcap G(n, \widetilde{p_n})$ on the probability space $\mathbb{S}$ such that $G_q(n, K_n, P_n) \bcap G(n, p_n)$ is a  spanning supergraph\footnote{A graph
$G_a$ is a spanning supergraph (resp., spanning subgraph) of a graph
$G_b$ if $G_a$ and $G_b$ have the same node set, and the edge set of
$G_a$ is a superset (resp., subset) of the edge set of $G_b$.\label{ft-span}} of $G_q(n, K_n, P_n) \bcap G(n, \widetilde{p_n})$   for all $n$ sufficiently large, where a sequence $\widetilde{\alpha_n}$   defined by
$s(K_n, P_n,q) \times \widetilde{p_n}=\frac{ \ln  n + {(k-1)} \ln \ln n +\widetilde{\alpha_n} }{n}$   satisfies ${\lim_{n \to \infty}{\widetilde{\alpha_n}} = \infty\text{ and }\widetilde{\alpha_n}= o(\ln n)}$\,.
\item[(ii)]
If ${\lim_{n \to \infty}\alpha_n   =  -\infty}$, there exists   a graph $G_q(n, \widehat{K_n}, P_n) \bcap G(n, \widehat{p_n})$ on the probability space $\mathbb{S}$ such that $G_q(n, K_n, P_n) \bcap G(n, p_n)$ is a spanning subgraph of   $G_q(n, \widehat{K_n}, P_n) \bcap G(n, \widehat{p_n})$ for all $n$ sufficiently large, where
\begin{align}
&\begin{array}{l} \textstyle{\frac{{\widehat{K_n}}^2}{P_n} = o\left( \frac{1}{\ln n} \right),  \frac{\widehat{K_n}}{P_n} = o\left( \frac{1}{n\ln n} \right) \text{and }} \\ \textstyle{\widehat{K_n}= \Omega(n^{\epsilon})\text{ for a positive constant }\epsilon} \end{array} \label{conditions-lem_Gq_cplinga2}
\end{align} and a sequence $\widehat{\alpha_n}$ defined by\\   $s(\widehat{K_n}, P_n,q) \times \widehat{p_n}=\frac{ \ln  n + {(k-1)} \ln \ln n +\widehat{\alpha_n} }{n}$  \vspace{1pt} satisfies  ${\lim_{n \to \infty}{\widehat{\alpha_n}} =- \infty\text{ and }\widehat{\alpha_n}= -o(\ln n)}$\,.

\end{itemize}
 }

\end{lem}

%

\noindent \textbf{Proof of (\ref{with_extra}) using Lemma \ref{lem_Gq_cplinga}:}

We now prove (\ref{with_extra}) using Lemma \ref{lem_Gq_cplinga}. Namely, assuming that
Theorem \ref{thm:OneLaw+NodeIsolation} holds with the confined deviation, we use   Lemma \ref{lem_Gq_cplinga} to  show that Theorem \ref{thm:OneLaw+NodeIsolation}  also holds regardless of the confined deviation.  To prove Theorem \ref{thm:OneLaw+NodeIsolation}, we discuss the two cases below:
\ding{172} $\lim_{n \to \infty}\alpha_n = \infty$, and \ding{173} $\lim_{n \to \infty}\alpha_n = -\infty$.

\ding{172} Under $\lim_{n \to \infty}\alpha_n = \infty$, we use the property (i) of Lemma \ref{lem_Gq_cplinga}, where we have  graph
$\mathbb{G}_{n,q}(n,{K_n},{P_n}, \widetilde{{p_n}})=G_q(n, K_n, P_n) \bcap G(n, \widetilde{p_n})$ with
$ K_n =
\Omega(n^{\epsilon})$ for a positive constant $\epsilon$, $ \frac{{K_n}^2}{P_n}  =
 o\left( \frac{1}{\ln n} \right)$,    $ \frac{K_n}{P_n} = o\left( \frac{1}{n\ln n} \right)$, and
$t( {K_n},{P_n}, q, \widetilde{{p_n}}) =  s(K_n, P_n,q) \times \widetilde{p_n}=\frac{\ln  n  + {(k-1)} \ln \ln n  +  {\widetilde{\alpha_n}}}{n}$. Then given $\lim_{n \to \infty}\widetilde{\alpha_n} = \infty$ and $\widetilde{\alpha_n} = o(\ln n)$, we use Theorem \ref{thm:OneLaw+NodeIsolation} with the confined deviation to derive
\begin{align}
\lim_{n \rightarrow \infty }  \mathbb{P} \left[\hspace{2pt}\mathbb{G}_{n,q}(n,K_n,P_n, \widetilde{{p_n}})\text{
is $k$-connected}.\hspace{2pt}\right] = 1. \label{zhx1}
 \end{align}
 As given in the property (i) of Lemma \ref{lem_Gq_cplinga}, $\mathbb{G}_{n,q}(n,\hspace{1pt}K_n,\hspace{1pt}P_n, \hspace{1pt}{p_n})$ is a spanning supergraph of $\mathbb{G}_{n,q}(n,\hspace{1pt}K_n,\hspace{1pt}P_n, \hspace{1pt}\widetilde{{p_n}})$.
 Then since $k$-connectivity is a monotone increasing graph property, we obtain from (\ref{zhx1}) that
  \begin{align}
& \mathbb{P} \left[\hspace{2pt} \mathbb{G}_{n,q}(n, K_n,P_n,{p_n})\text{
is $k$-connected}.\hspace{2pt}\right]  &
 \nonumber \\
  &   \geq   \mathbb{P} \left[\hspace{2pt}\mathbb{G}_{n,q}(n,{K_n},{P_n}, \widetilde{{p_n}})\text{
is $k$-connected}.\hspace{2pt}\right]
  \to 1. \text{ as $n \rightarrow \infty$.}
  \label{graph_g_a_g_b}
\end{align}
(\ref{graph_g_a_g_b}) provides the desired result (\ref{thm-con-eq-1}).

\ding{173} Under $\lim_{n \to \infty}\alpha_n = -\infty$, we use the property (ii) of Lemma \ref{lem_Gq_cplinga}, where we have  graph
$\mathbb{G}_{n,q}(n,\widehat{K_n},{P_n}, \widehat{{p_n}})=G_q(n, \widehat{K_n}, P_n) \bcap G(n, \widehat{p_n})$ with$\widehat{K_n}= \Omega(n^{\epsilon})$ for a positive constant $\epsilon$,
$\frac{{\widehat{K_n}}^2}{P_n} = o\left( \frac{1}{\ln n} \right)$,  $\frac{\widehat{K_n}}{P_n} = o\left( \frac{1}{n\ln n} \right) $, and
$t( \widehat{K_n},{P_n}, q, \widehat{{p_n}}) = s(\widehat{K_n}, P_n,q) \times \widehat{p_n}= \frac{\ln  n  + {(k-1)} \ln \ln n  +  {\widehat{\alpha_n}}}{n}$. Then given $\lim_{n \to \infty}\widehat{\alpha_n} = -\infty$ and $\widehat{\alpha_n} =- o(\ln n)$, we use Theorem \ref{thm:OneLaw+NodeIsolation} with the confined deviation to derive
\begin{align}
&\lim_{n \rightarrow \infty }  \mathbb{P} \big[\hspace{2pt}\mathbb{G}_{n,q}(n,\widehat{K_n},{P_n}, \widehat{{p_n}})\text{
is $k$-connected}.\hspace{2pt}\big] = 0. \label{one-zhx1}
 \end{align}
 As given in the property (ii) of Lemma \ref{lem_Gq_cplinga}, $\mathbb{G}_{n,q}(n, K_n,P_n, {p_n})$ is a spanning subgraph of $\mathbb{G}_{n,q}(n,\widehat{K_n},P_n, \widehat{{p_n}})$.
  Then since $k$-connectivity is a monotone increasing graph property, we obtain from (\ref{one-zhx1}) that
  \begin{align}
& \mathbb{P} \left[\hspace{2pt} \mathbb{G}_{n,q}(n, K_n,P_n,{p_n})\text{
is $k$-connected}.\hspace{2pt}\right]  &
 \nonumber \\
  &   \leq   \mathbb{P} \big[\hspace{2pt}\mathbb{G}_{n,q}(n,\widehat{K_n},{P_n}, \widehat{{p_n}})\text{
is $k$-connected}.\hspace{2pt}\big]
  \to 0. \text{ as $n \rightarrow \infty$.}
  \label{one-graph_g_a_g_b}
\end{align}
(\ref{one-graph_g_a_g_b}) provides the desired result  (\ref{thm-con-eq-0}).

Summarizing \ding{172} and \ding{173}, we have established (\ref{with_extra}).
 Hence, in proving Theorem \ref{thm:OneLaw+NodeIsolation}, we can always assume $|{\alpha_n} |=  o ( \ln n)$.

\noindent \textbf{Proof of Lemma \ref{lem_Gq_cplinga}:}

 We prove Properties (i) and (ii) of Lemma \ref{lem_Gq_cplinga}, respectively.

\textbf{Establishing Property (i) of Lemma \ref{lem_Gq_cplinga}:}

We define
\begin{align}
\widetilde{\alpha_n} = \min\{\alpha_n,~\ln\ln n\}, \label{widetilde-alpha-n-def}
\end{align}
and define $\widetilde{p_n}$ such that
\begin{align}
s(K_n, P_n,q) \times \widetilde{p_n}=\frac{ \ln  n + {(k-1)} \ln \ln n +\widetilde{\alpha_n} }{n}. \label{widetilde-alpha-n-def2}
\end{align}


 Given the condition $\lim_{n \to \infty}\alpha_n = \infty$ in Property (i) of Lemma \ref{lem_Gq_cplinga}, we have $\alpha_n \geq 0$ for all $n$ sufficiently large, which with (\ref{widetilde-alpha-n-def}) implies
  \begin{align}
\text{$0 \leq \widetilde{\alpha_n}  \leq \ln\ln n$ for all $n$ sufficiently large}. \label{widetilde-alpha-n-def4a}
\end{align}
 Thus, it holds that
 \begin{align}
\widetilde{\alpha_n}= o(\ln n). \label{widetilde-alpha-n-def4}
\end{align}
 In addition, $\lim_{n \to \infty}\alpha_n = \infty$ and (\ref{widetilde-alpha-n-def}) together induce
  \begin{align}
\lim_{n \to \infty}{\widetilde{\alpha_n}} = \infty. \label{widetilde-alpha-n-def5}
\end{align}

Clearly, (\ref{widetilde-alpha-n-def}) implies $\widetilde{\alpha_n} \leq \alpha_n$.
Given $\widetilde{\alpha_n} \leq \alpha_n$, (\ref{widetilde-alpha-n-def2}) and
$s(K_n, P_n,q) \times p_n  = \frac{\ln  n + {(k-1)} \ln \ln n + {\alpha_n}}{n}$, we obtain $\widetilde{p_n} \leq p_n$. In addition, we know from (\ref{widetilde-alpha-n-def2}) and (\ref{widetilde-alpha-n-def4a}) that $\widetilde{p_n} \geq 0$ for all $n$ sufficiently large. For all $n$ sufficiently large, given $0 \leq \widetilde{p_n}\leq p_n \leq 1$,  $\widetilde{p_n}$ is indeed a probability, and we can define Erd\H{o}s--R\'{e}nyi graphs $G(n,p_n)$ and  $G(n,\widetilde{p_n})$ on the same probability space  such that $G(n, p_n)$ is a spanning supergraph of $G(n, \widetilde{p_n})$. Then we can define $G_q(n, K_n, P_n) \bcap G(n, p_n)$ and  $G_q(n, K_n, P_n) \bcap G(n, \widetilde{p_n})$
on the same probability space  such that
\begin{align}
\begin{array}{l} \text{\textit{$G_q(n, K_n, P_n) \bcap G(n, p_n)$ is a spanning supergraph of}} \\ \text{\textit{$G_q(n, K_n, P_n) \bcap G(n, \widetilde{p_n})$.}}  \end{array} \label{widetilde-alpha-n-def3}
\end{align}

Summarizing (\ref{widetilde-alpha-n-def4}) (\ref{widetilde-alpha-n-def5}) and (\ref{widetilde-alpha-n-def3}), we have established Lemma \ref{lem_Gq_cplinga}.

\textbf{Establishing Property (ii) of Lemma \ref{lem_Gq_cplinga}:}

To establish Property (ii) of Lemma \ref{lem_Gq_cplinga}, we may attempt to use a proof similar to that of  Property (i) of Lemma \ref{lem_Gq_cplinga}, by defining $\widehat{\alpha_n}$ as $\max\{\alpha_n,~-\ln\ln n\}$,
and defining $\widehat{p_n}$ such that
$s(K_n, P_n,q) \times \widehat{p_n}$ equals $\frac{ \ln  n + {(k-1)} \ln \ln n +\widehat{\alpha_n} }{n}$. However, {{such approach does not work}} because $\widehat{p_n}$ defined in this way may exceed $1$ so it is not a probability. Hence, more fine-grained arguments are needed. In view of the above, we consider two cases for each $n$:
\begin{itemize}
\item[\ding{202}]
  $s(K_n, P_n,q)\geq\frac{\ln  n + {(k-1)} \ln \ln n + \max\{\alpha_n,-\ln\ln n\}}{n}$,
\item[\ding{203}]
$s(K_n, P_n,q)<\frac{\ln  n + {(k-1)} \ln \ln n + { \max\{\alpha_n,-\ln\ln n\}}}{n}$.
\end{itemize}
 In the above case \ding{202}, we can define $\widehat{p_n}$ in the above way since we can show $\widehat{p_n}\leq 1$ for all $n$ sufficiently large. In the above case \ding{203}, since $\widehat{p_n}$ defined in the above way may exceed $1$, we will define $\widehat{p_n}$ differently. More specifically, in case \ding{203}, we will find suitable   $\widehat{p_n} \geq p_n$ and $\widehat{K_n} \geq K_n$ such that
$s(\widehat{K_n}, P_n,q) \times \widehat{p_n}$ equals   $\frac{ \ln  n + {(k-1)} \ln \ln n +\widehat{\alpha_n} }{n}$ for some $\widehat{\alpha_n} $ satisfying $\lim_{n \to \infty}{\widehat{\alpha_n}} =- \infty\text{ and }|\widehat{\alpha_n}|= o(\ln n)$. We will carefully choose the term ${\widehat{\alpha_n}}$ in case \ding{203}  rather than simply setting ${\widehat{\alpha_n}}$ as $\max\{\alpha_n,~-\ln\ln n\}$. We provide the details below.


\begin{itemize}[leftmargin=3pt]
\item[\ding{202}]
In this case, we consider
\begin{align}
s(K_n, P_n,q)\geq\frac{\ln  n + {(k-1)} \ln \ln n + \max\{\alpha_n,-\ln\ln n\}}{n}. \label{widehat-alpha-n-def}
\end{align}
Then we define
\begin{align}
\widehat{K_n}&=K_n  \text{ in case \ding{202}}, \label{widehat-alpha-n-def2} \\
\widehat{\alpha_n}&=\max\{\alpha_n,-\ln\ln n\} \text{ in case \ding{202}},\label{widehat-alpha-n-def3}
\end{align}
and define $\widehat{p_n}$ such that
\begin{align}
\widehat{p_n} \cdot s(K_n, P_n,q) = \frac{\ln  n + {(k-1)} \ln \ln n + {\widehat{\alpha_n}}}{n} \text{ in case \ding{202}}. \label{widehat-alpha-n-def4}
\end{align}
From (\ref{widehat-alpha-n-def4}) and the condition (\ref{widehat-alpha-n-def}) in case \ding{202} here, we have
\begin{align}
\widehat{p_n} \leq 1 \text{ in case \ding{202}}. \label{widehat-alpha-n-def4b}
\end{align}

Clearly, (\ref{widehat-alpha-n-def3}) implies $\widehat{\alpha_n} \geq \alpha_n$.
Given $\widehat{\alpha_n} \geq \alpha_n$, (\ref{widehat-alpha-n-def4}) and
$s(K_n, P_n,q) \times p_n  = \frac{\ln  n + {(k-1)} \ln \ln n + {\alpha_n}}{n}$, we obtain
  \begin{align}
\widehat{p_n} \geq p_n \text{ in case \ding{202}}. \label{widehat-alpha-n-def5a}
\end{align}

 Given the condition $\lim_{n \to \infty}\alpha_n = -\infty$  in Property (ii) of Lemma \ref{lem_Gq_cplinga}, we have $\alpha_n \leq 0$ for all $n$ sufficiently large, which with (\ref{widehat-alpha-n-def3}) implies
  \begin{align}
\text{$-\ln\ln n\leq \widehat{\alpha_n}  \leq 0 $ for all $n$ sufficiently large} \text{ in case \ding{202}}. \label{widehat-alpha-n-def5}
\end{align}
\item[\ding{203}]
In this case, we consider
\begin{align}
s(K_n, P_n,q)<\frac{\ln  n + {(k-1)} \ln \ln n + { \max\{\alpha_n,-\ln\ln n\}}}{n}. \label{widehat-alpha-n-defz}
\end{align}
Then we define
\begin{align}
\widehat{p_n}&=1 \text{ in case \ding{203}} , \label{widehat-alpha-n-defz2}
\end{align}
define that
\begin{align}
\begin{array}{l} \text{\textit{\text{in case \ding{203}}, $\widehat{K_n}$ is the maximal integer $K_n^{\#}$ such that}} \\ \text{\textit{$ s_{n,q}(K_n^{\#}, P_n) $ is no greater than}} \\ \text{\textit{$ \frac{\ln  n + {(k-1)} \ln \ln n + { \max\{\alpha_n,-\ln\ln n\}}}{n}$,}}  \end{array} \label{widehat-alpha-n-defz3}
\end{align}
and define $\widehat{\alpha_n}$ such that
\begin{align}
 s(\widehat{K_n}, P_n,q) = \frac{\ln  n + {(k-1)} \ln \ln n + {\widehat{\alpha_n}}}{n} \text{ in case \ding{203}}. \label{widehat-alpha-n-defz4}
\end{align}

From (\ref{widehat-alpha-n-defz2}) and $p_n \leq 1$ since $p_n$ is a probability, it holds that
\begin{align}
\widehat{p_n} \geq p_n \text{ in case \ding{203}}. \label{widehat-alpha-n-defz5}
\end{align}
From (\ref{widehat-alpha-n-defz}) and (\ref{widehat-alpha-n-defz3}), it holds that
\begin{align}
\widehat{K_n} \geq K_n \text{ in case \ding{203}}. \label{widehat-alpha-n-defz6}
\end{align}
 \end{itemize}

Combining (\ref{widehat-alpha-n-def2}) for case \ding{202} and (\ref{widehat-alpha-n-defz6}) for case \ding{203}, we have
\begin{align}
\widehat{K_n} \geq K_n \text{ for all $n$}. \label{widehat-alpha-n-defz8}
\end{align}
From (\ref{widehat-alpha-n-defz8}) and the condition $K_n =
\omega(1)$ of Lemma \ref{lem_Gq_cplinga}-Property (ii) here, we have
\begin{align}
\widehat{K_n} = \omega(1). \label{widehat-alpha-n-defz8ac}
\end{align}
Combining (\ref{widehat-alpha-n-def5a}) for case \ding{202} and (\ref{widehat-alpha-n-defz5}) for case \ding{203}, we have
\begin{align}
\widehat{p_n} \geq p_n \text{ for all $n$}. \label{widehat-alpha-n-defz9}
\end{align} Combining (\ref{widehat-alpha-n-def4b}) for case \ding{202} and (\ref{widehat-alpha-n-defz2}) for case \ding{203}, we have
\begin{align}
\widehat{p_n} \leq 1 \text{ for all $n$}. \label{widehat-alpha-n-defz10}
\end{align}
 Then given (\ref{widehat-alpha-n-defz8}) (i.e., $\widehat{K_n} \geq K_n$  for each $n$), from the definitions of graphs $G_q(n, K_n, P_n)$ and  $G_q(n, \widehat{K_n}, P_n)$, we can construct them on the same probability space  such that $G_q(n, K_n, P_n)$ is a spanning subgraph of $G_q(n, \widehat{K_n}, P_n)$. Given (\ref{widehat-alpha-n-defz9}) and (\ref{widehat-alpha-n-defz10}) (i.e., $p_n \leq \widehat{p_n} \leq 1$  for each $n$), $\widehat{p_n}$ is indeed a probability, and  we can define Erd\H{o}s--R\'{e}nyi graphs $G(n,p_n)$ and  $G(n,\widehat{p_n})$ on the same probability space  such that $G(n, p_n)$ is a spanning subgraph of $G(n, \widehat{p_n})$. Summarizing the above, we can define $G_q(n, K_n, P_n) \bcap G(n, p_n)$ and  $G_q(n, \widehat{K_n}, P_n) \bcap G(n, \widehat{p_n})$
on the same probability space  such that
\begin{align}
\begin{array}{l} \text{\textit{$G_q(n, K_n, P_n) \bcap G(n, p_n)$ is a spanning subgraph of}} \\ \text{\textit{$G_q(n, \widehat{K_n}, P_n) \bcap G(n, \widehat{p_n})$.}}  \end{array} \label{widehat-alpha-n-defz7}
\end{align}
Given (\ref{widehat-alpha-n-defz7}), we now show the results on $\widehat{K_n}$ and $\widehat{\alpha_n}$ to complete the proof of
Lemma \ref{lem_Gq_cplinga}-Property (ii).

From the condition $\frac{{K_n}^2}{P_n} = o(1)$ of Lemma \ref{lem_Gq_cplinga} here, we have $K_n < P_n$ for all $n$ sufficiently large. Then from (\ref{widehat-alpha-n-def2}), we get
\begin{align}
\widehat{K_n}&< P_n \text{ for all $n$ sufficiently large, in case \ding{202}}, \nonumber
\end{align}
so that we can evaluate $ s_{n,q}(\widehat{K_n} + 1, P_n) $ for all $n$ sufficiently large, in case \ding{202} here.
From $ s_{n,q}(\widehat{K_n} + 1, P_n) \geq  s(K_n, P_n,q)$ and (\ref{widehat-alpha-n-def}), it follows that
\begin{align}
&s_{n,q}(\widehat{K_n} + 1, P_n)\nonumber \\ &\geq\frac{\ln  n + {(k-1)} \ln \ln n + \max\{\alpha_n,-\ln\ln n\}}{n}\nonumber \\ &\text{for all $n$ sufficiently large, in case \ding{202}}. \label{widehat-alpha-n-def2-cnt4}
\end{align}

Clearly, it holds that
$\frac{\ln  n + {(k-1)} \ln \ln n + { \max\{\alpha_n,-\ln\ln n\}}}{n} < 1$ for all $n$ sufficiently large. Given this, (\ref{widehat-alpha-n-defz3}), and $s_{n,q}(P_n, P_n) = 1$, we obtain
\begin{align}
\widehat{K_n}&< P_n \text{ for all $n$ sufficiently large, in case \ding{203}}\nonumber
\end{align}
so that we can evaluate $ s_{n,q}(\widehat{K_n} + 1, P_n) $ for all $n$ sufficiently large, in case \ding{202} here. Then (\ref{widehat-alpha-n-defz3}) implies
\begin{align}
&s_{n,q}(\widehat{K_n} + 1, P_n)\nonumber \\ &>\frac{\ln  n + {(k-1)} \ln \ln n + \max\{\alpha_n,-\ln\ln n\}}{n})\nonumber \\ &\text{for all $n$ sufficiently large, in case \ding{203}}. \label{widehat-alpha-n-def2-cnt5}
\end{align}

Combining (\ref{widehat-alpha-n-def2-cnt4}) and (\ref{widehat-alpha-n-def2-cnt5}), we have
\begin{align}
&s_{n,q}(\widehat{K_n} + 1, P_n)\nonumber \\ &\geq\frac{\ln  n + {(k-1)} \ln \ln n + \max\{\alpha_n,-\ln\ln n\}}{n}\nonumber \\ &\text{for all $n$ sufficiently large}. \label{widehat-alpha-n-def2-cnt6}
\end{align}

From (\ref{widehat-alpha-n-def}), it follows that
\begin{align}
&s(\widehat{K_n}, P_n,q) \nonumber \\ &\leq \max\left\{\begin{array}{l}s(K_n, P_n,q), \\[3pt]  \frac{\ln  n + {(k-1)} \ln \ln n + { \max\{\alpha_n,-\ln\ln n\}}}{n} \end{array}\right\} \text{ for all $n$},\nonumber
\end{align}
which implies
\begin{align}
&s(\widehat{K_n}, P_n,q) = o(1).\label{widehat-alpha-n-def2-cnt11}
\end{align}

Given (\ref{widehat-alpha-n-defz8ac}) and (\ref{widehat-alpha-n-def2-cnt11}), we use Lemma \ref{lem_eval_psq}-Property (i) to obtain
\begin{align}
&\frac{{\widehat{K_n}}^2}{P_n} = o(1).\label{widehat-alpha-n-def2-cnt12}
\end{align}
Given (\ref{widehat-alpha-n-defz8ac}) and (\ref{widehat-alpha-n-def2-cnt12}), we use Lemma \ref{lem_eval_psq}-Property (i) to obtain
   \begin{align}
s(\widehat{K_n}, P_n,q)&  = \frac{1}{q!} \bigg(\frac{{\widehat{K_n}}^2}{P_n}\bigg)^{q} \times [1\pm o(1)].\label{widehat-alpha-n-def2-cnt13}
 \end{align}
 Given (\ref{widehat-alpha-n-defz8ac}) and (\ref{widehat-alpha-n-def2-cnt12}), we also have $\widehat{K_n} + 1 = \omega(1)$ and $\frac{{(\widehat{K_n} + 1)}^2}{P_n} = o(1)$. Then we use Lemma \ref{lem_eval_psq}-Property (i) to obtain
 \begin{align}
& s_{n,q}(\widehat{K_n} + 1, P_n)  = \frac{1}{q!} \bigg(\frac{{(\widehat{K_n} + 1)}^2}{P_n}\bigg)^{q} \times [1\pm o(1)].\label{widehat-alpha-n-def2-cnt14}
 \end{align}
From (\ref{widehat-alpha-n-def2-cnt13}) (\ref{widehat-alpha-n-def2-cnt14}) and (\ref{widehat-alpha-n-defz8ac}) , it follows that
 \begin{align}
\frac{s_{n,q}(\widehat{K_n} + 1, P_n)}{s(\widehat{K_n}, P_n,q)} &  \sim \frac{{(\widehat{K_n} + 1)}^2}{P_n}  \bigg/  \frac{{\widehat{K_n}}^2}{P_n} \nonumber \\ &  = \bigg(1+\frac{1}{\widehat{K_n}}\bigg)^{2} \to 1,\text{ as }n \to \infty ,\label{widehat-alpha-n-def2-cnt15}
 \end{align}
where the expression $a_n \sim b_n$ for two positive sequences $a_n$ and $b_n$ means $\lim_{n \to \infty} ( {a_n}/{b_n})=1$.

Combining (\ref{widehat-alpha-n-def2-cnt6}) and (\ref{widehat-alpha-n-def2-cnt15}), we have
\begin{align}
&s(\widehat{K_n}, P_n,q)\nonumber \\ &\geq\frac{\ln  n + {(k-1)} \ln \ln n + \max\{\alpha_n,-\ln\ln n\}}{n} \times [1-o(1)]\nonumber \\ &= \frac{\ln  n + {(k-1)} \ln \ln n + \max\{\alpha_n,-\ln\ln n\}-o(\ln n)}{n}  \label{widehat-alpha-n-def2-cnt6at} ,
\end{align}
where the last step uses
\begin{align}
\max\{\alpha_n,-\ln\ln n\} = -o(\ln n) . \label{widehat-alpha-n-defz4acr6}
\end{align}
The result
(\ref{widehat-alpha-n-defz4acr6}) follows because we have $-\ln\ln n \leq \max\{\alpha_n,-\ln\ln n\} <0$ given $\alpha_n <0$ for all $n$ sufficiently large from the condition $\lim_{n \to \infty}\alpha_n   =  -\infty$ of Lemma \ref{lem_Gq_cplinga}-Property (ii) here.

Then (\ref{widehat-alpha-n-def2-cnt6at}) means that $\alpha_n^{\#}$ defined by
\begin{align}
 s(\widehat{K_n}, P_n,q) = \frac{\ln  n + {(k-1)} \ln \ln n + \alpha_n^{\#}}{n}  \label{widehat-alpha-n-defz4acr}
\end{align}
satisfies
\begin{align}
\alpha_n^{\#} \geq -o(\ln n) . \label{widehat-alpha-n-defz4acr2}
\end{align}
From (\ref{widehat-alpha-n-def3}) (\ref{widehat-alpha-n-defz4}) and (\ref{widehat-alpha-n-defz4acr}),  we have\\ $\widehat{\alpha_n} = \begin{cases} \max\{\alpha_n,-\ln\ln n\}&\text{in case \ding{202}}, \\
\alpha_n^{\#} &\text{in case \ding{203}} . \end{cases}$ Then it holds that
\begin{align}
\widehat{\alpha_n} \geq \min\{\max\{\alpha_n,-\ln\ln n\}, \alpha_n^{\#}\}  \geq -o(\ln n)  \label{widehat-alpha-n-defz4acr3} ,
\end{align}
where the last step uses (\ref{widehat-alpha-n-defz4acr6}) and (\ref{widehat-alpha-n-defz4acr2}).

From (\ref{widehat-alpha-n-def3}) (\ref{widehat-alpha-n-defz}) and (\ref{widehat-alpha-n-defz4}),  we have
\begin{align}
\widehat{\alpha_n} \leq \max\{\alpha_n,-\ln\ln n\}, \label{widehat-alpha-n-defz4acr5cd}
\end{align}
 which along with (\ref{widehat-alpha-n-defz4acr6}) will imply
\begin{align}
\widehat{\alpha_n} \leq -o(\ln n).  \label{widehat-alpha-n-defz4acr4}
\end{align}
Combining (\ref{widehat-alpha-n-defz4acr3}) and (\ref{widehat-alpha-n-defz4acr4}),  we have
\begin{align}
\widehat{\alpha_n} = -o(\ln n) . \label{widehat-alpha-n-defz4acr5}
\end{align}
From (\ref{widehat-alpha-n-defz4acr5cd}) and the condition $\lim_{n \to \infty}\alpha_n   =  -\infty$ of Lemma \ref{lem_Gq_cplinga}-Property (ii), it holds that
\begin{align}
\lim_{n \to \infty}\widehat{\alpha_n}  =  -\infty .  \label{widehat-alpha-n-defz4acr5acrd}
\end{align}

Summarizing
(\ref{widehat-alpha-n-defz8ac}) (\ref{widehat-alpha-n-def2-cnt12}) (\ref{widehat-alpha-n-defz7}) (\ref{widehat-alpha-n-defz4acr5}) and (\ref{widehat-alpha-n-defz4acr5acrd}), we have completed showing Lemma \ref{lem_Gq_cplinga}-Property (ii).

Given the above, we have proved both properties of Lemma~\ref{lem_Gq_cplinga}. \qeda

\begin{lem} \label{lem_eval_psq}
The following two properties hold, where $s_{n,q} $ denotes the probability that two  nodes in graph $\mathbb{G}_q$
share at least $q$ keys:
\begin{itemize}
\item[(i)] If $K_n = \omega(1)$ and $\frac{{K_n}^2}{P_n} = o(1)$, then\\$s_{n,q}
= \frac{1}{q!} \big( \frac{{K_n}^2}{P_n} \big)^{q} \times [1\pm o(1)]$; i.e., $s_{n,q}
\sim \frac{1}{q!} \big( \frac{{K_n}^2}{P_n} \big)^{q}$.\vspace{3pt}
\item[(ii)] If $K_n = \omega(\ln n)$ and $\frac{{K_n}^2}{P_n} = o\big(\frac{1}{\ln n}\big)$, then\\$s_{n,q}
= \frac{1}{q!} \big( \frac{{K_n}^2}{P_n} \big)^{q} \times [1\pm o\big(\frac{1}{\ln n}\big)]$.
\end{itemize}
\end{lem}

Lemma \ref{lem_eval_psq} can be proved in a way similar to that of \cite[Lemma 1]{zhao2014topological}. We omit the details due to space limitation.

\section{Proving the Lower Bound of   Section \ref{sec-Decomposing}} \label{sec-prf-lower-bounds-details}

The idea to prove the lower bound $e^{-
\frac{e^{- \iffalse \lim\limits_ \fi \lim_{n \to \infty}{\alpha_n}}}{(k-1)!}} \times [1 - o(1)]$ for $\mathbb{P}[\hspace{2pt}\mathbb{G}_{n,q}\text{ is
  $k$-connected.}\hspace{2pt}]$ has been explained in Section \ref{sec-prf-lower-bounds}. As explained, we associate the studied graph $\mathbb{G}_{n,q}$ with an Erd\H{o}s--R\'enyi graph $G(n,z_n)$. The result is presented as Lemma \ref{lem-cpgraph-rigrig} below.

Lemma \ref{lem-cp_rig_er} relates our graph   $\mathbb{G}_{n,q}$ with an Erd\H{o}s--R\'enyi graph.

\begin{lem} \label{lem-cp_rig_er}
 If $ K_n =
\Omega(n^{\epsilon})$ for a positive constant $\epsilon$, \vspace{1pt} $ \frac{{K_n}^2}{P_n}  =
 o\left( \frac{1}{\ln n} \right)$, $ \frac{K_n}{P_n} = o\left( \frac{1}{n\ln n} \right)$, and $\frac{{K_n}^2}{P_n} = \omega\big(\frac{(\ln n)^6}{n^2}\big)$, then there exists a sequence $z_n$ satisfying
\begin{align}
\textstyle{z_n = t_{n,q} \times \big[1-o\big(\frac{1}{\ln n}\big)\big]} \label{ERgraph-sn-defn}
\end{align}
 such that
graph $\mathbb{G}_{n,q}$ contains an Erd\H{o}s--R\'{e}nyi graph $G(n,z_n)$ as a spanning subgraph with probability $1-o(1)$ (when we couple the two graphs on the same probability space and define them on the same node set), where we note that $t_{n,q}$ is the edge probability of $\mathbb{G}_{n,q}$, and $z_n$ is the edge probability of $G(n,z_n)$.
 \end{lem}

\begin{rem}
From \cite{zz}, since $k$-connectivity is a monotone increasing graph property, (\ref{ERgraph-sn-defn}) further implies
\begin{align}
 \bP{\text{$\mathbb{G}_{n,q}\iffalse_{on}^{(q)}\fi$  is $k$-connected.}}   \geq \bP{\text{$G(n,z_n)$ is $k$-connected.}} - o(1) . \label{Gq-kcon-lower-bound-tonsb}
\end{align}
 \end{rem}

Recall from (\ref{eq:G_on_is_RKG_cap_ER_oy}) that $\mathbb{G}_{n,q}$ is the intersection of a uniform $q$-intersection graph $G_q(n, K_n,P_n)$ and an
Erd\H{o}s-R\'enyi graph $G(n,{p_n\iffalse_{on}\fi})$. To prove Lemma \ref{lem-cp_rig_er} which associates $\mathbb{G}_{n,q}$ with an
Erd\H{o}s-R\'enyi graph, we   establish Lemma \ref{lem-cpgraph-rigrig} below which associates $G_q(n, K_n,P_n)$ with another
Erd\H{o}s-R\'enyi graph.

\begin{lem} \label{lem-cpgraph-rigrig}
 If $ K_n =
\Omega(n^{\epsilon})$ for a positive constant $\epsilon$, \vspace{1pt} $ \frac{{K_n}^2}{P_n}  =
 o\left( \frac{1}{\ln n} \right)$, $ \frac{K_n}{P_n} = o\left( \frac{1}{n\ln n} \right)$, and $\frac{{K_n}^2}{P_n} = \omega\big(\frac{(\ln n)^6}{n^2}\big)$, then there exists a sequence $y_n$ satisfying
\begin{align}
\textstyle{y_n = s_{n,q} \times \big[1-o\big(\frac{1}{\ln n}\big)\big]} \label{ERgraph-sn-defn-reduced}
\end{align}
such that
a uniform $q$-intersection graph $G_q(n, K_n,P_n)$ contains an Erd\H{o}s--R\'{e}nyi graph $G(n,y_n)$ as a spanning subgraph with probability $1-o(1)$ (when we couple the two graphs on the same probability space and define them on the same node set), where $s_{n,q}$ is the edge probability of $G_q(n, K_n, P_n)$.
 \end{lem}

We will discuss the proof of Lemma \ref{lem-cpgraph-rigrig} later. Below we show that Lemma \ref{lem-cp_rig_er} follows from Lemma \ref{lem-cpgraph-rigrig}.

\textbf{Proof of Lemma \ref{lem-cp_rig_er} using Lemma \ref{lem-cpgraph-rigrig}:}

As noted in Lemmas  \ref{lem-cp_rig_er} and \ref{lem-cpgraph-rigrig}, we will couple different random graphs together. The goal is to convert a problem in one random graph to the corresponding problem in another random graph, in order to solve the original problem.
Formally, a coupling
\cite{zz,2013arXiv1301.0466R,Krzywdzi} of two random graphs
$G_1$ and $G_2$ means a probability space on which random graphs
$G_1'$ and $G_2'$ are defined such that $G_1'$ and $G_2'$ have the
same distributions as $G_1$ and $G_2$, respectively. For notation
brevity, we simply say $G_1$ is a spanning subgraph
(resp., spanning supergraph) of $G_2$ if $G_1'$ is a spanning
subgraph, where the notions of spanning subgraph and supergraph have been defined in Footnote (\ref{ft-span}).

 Following Rybarczyk's notation \cite{zz}, we
write
\begin{align}
G_1 \succeq & G_2 \quad (\textrm{resp.}, G_1 \succeq_{1-o(1)} G_2)
\label{g1g2coupling}
\end{align}
if there exists a coupling under which $G_2$ is a spanning subgraph
of $G_1$ with probability $1$ (resp., $1-o(1)$); i.e., $G_1$ is a spanning supergraph
of $G_2$ with probability $1$ (resp., $1-o(1)$).
Then the conclusion in Lemma \ref{lem-cp_rig_er} means
\begin{align}
\mathbb{G}_{n,q}  \succeq_{1-o(1)}G(n,z_n),  \label{lem1-rescp}
\end{align}
while the conclusion in Lemma \ref{lem-cpgraph-rigrig} means
\begin{align}
G_q(n,K_n,P_n)   \succeq_{1-o(1)}G(n,y_n). \label{lem1-rescp}
\end{align}

We recall from (\ref{eq:G_on_is_RKG_cap_ER_oy}) that
\begin{equation}
\mathbb{G}_{n,q}= G_q(n, K_n, P_n) \cap G(n, p_n).
 \label{eq:G_on_is_RKG_cap_ER_oyton-reduced}
\end{equation}
After intersecting $G_q(n, K_n, P_n)$ (resp., $G(n,y_n)$) with $G(n, p_n)$, we obtain $G_q(n, K_n, P_n) \cap G(n, p_n)$ (resp., $G(n,y_n) \cap G(n, p_n)$), where $G_q(n, K_n, P_n) \cap G(n, p_n)$ is $\mathbb{G}_{n,q}$ from (\ref{eq:G_on_is_RKG_cap_ER_oyton-reduced}), and $G(n,y_n) \cap G(n, p_n)$ becomes an Erd\H{o}s--R\'{e}nyi graph $G(n,y_n p_n)$. From (\ref{lem1-rescp}) (i.e., Lemma \ref{lem-cpgraph-rigrig}), $G_q(n, K_n, P_n)$ contains an Erd\H{o}s--R\'{e}nyi graph $G(n,y_n)$ as a spanning subgraph with probability $1-o(1)$ for $y_n$ in (\ref{ERgraph-sn-defn-reduced}) (when we couple the two graph intersections on the same probability space and define them on the same node set). Then $\mathbb{G}_{n,q}$ contains an Erd\H{o}s--R\'{e}nyi graph $G(n,y_n p_n)$ as a spanning subgraph with probability $1-o(1)$ for $y_n$ in (\ref{ERgraph-sn-defn-reduced}) (when we couple the two graph intersections on the same probability space and define them on the same node set); i.e.,
\begin{align}
\mathbb{G}_{n,q} \succeq_{1-o(1)} G(n,y_n p_n).
\label{g1g2coupling-v2}
\end{align}  Hence, the proof of Lemma \ref{lem-cp_rig_er} will be completed once we show $z_n$ in (\ref{ERgraph-sn-defn}) can be set as $y_n p_n$. From (\ref{ERgraph-sn-defn}) and $t_{n,q} = s_{n,q} p_n$, it follows that
\begin{align}
y_n p_n & = s_{n,q} \times \textstyle{\big[1-o\big(\frac{1}{\ln n}\big)\big]}  \times  p_n =   t_{n,q} \times \textstyle{\big[1-o\big(\frac{1}{\ln n}\big)\big]}. \nonumber
\end{align}
Hence, $z_n$ in (\ref{ERgraph-sn-defn}) can be set as $y_n p_n$. Then as explained above, we have proved Lemma \ref{lem-cp_rig_er} using Lemma \ref{lem-cpgraph-rigrig}. \qeda

\textbf{Basic Ideas of Proving Lemma \ref{lem-cpgraph-rigrig}:}

We now discuss the proof of Lemma \ref{lem-cpgraph-rigrig}.
The proof of Lemma \ref{lem-cpgraph-rigrig} is quite involved, since uniform $q$-intersection graph $G_q(n, K_n,P_n)$ and Erd\H{o}s--R\'{e}nyi graph $G(n,y_n)$ associated by Lemma \ref{lem-cpgraph-rigrig} are very different. For instance, while edges in $G(n,y_n)$ are all independent, not all edges in $G_q(n, K_n,P_n)$ are  independent with each other, since the event that nodes $v_1$ and $v_2$ share at least $q$ objects, and the event that nodes $v_1$ and $v_3$ share at least $q$ objects, may induce higher chance for  the event that nodes $v_2$ and $v_3$ share at least $q$ objects.

 To prove Lemma \ref{lem-cpgraph-rigrig}, we
introduce an auxiliary graph called the \emph{binomial
$q$-intersection
 graph} $H_q(n,x_n,P_n)$ \cite{Rybarczyk,Assortativity,bloznelis2013}, which
 can be defined on $n$ nodes by the following process.
  There exists an object pool of size $P_n$. Each object in the
pool is added to each node {independently} with probability
$x_n$. After each node obtains a set of objects, two nodes establish an edge in between if and only if they share at least $q$ objects. Clearly, the only difference between binomial $q$-intersection
 graph $H_q(n,x_n,P_n)$ and uniform $q$-intersection   graph $G_q(n,K_n,P_n)$ is that in the former,
  the number of objects assigned to each
 node obeys a binomial distribution with $P_n$ as
the number of trials, and with $x_n$ as the success probability in
each trial, while in the latter graph, such number equals $K_n$ with
probability $1$.

To prove Lemma \ref{lem-cpgraph-rigrig}, we present Lemmas \ref{brig_urig} and \ref{er_brig} below. Lemma \ref{brig_urig} shows that a uniform $q$-intersection graph $G_q(n, K_n,P_n)$ contains a binomial $q$-intersection
 graph $H_q(n,x_n,P_n)$ as a spanning subgraph with probability $1-o(1)$ (when we couple the two graphs on the same probability space and define them on the same node set). Lemma \ref{er_brig} shows that a binomial $q$-intersection
 graph $H_q(n,x_n,P_n)$ contains an Erd\H{o}s--R\'{e}nyi graph $G(n,y_n)$ as a spanning subgraph with probability $1-o(1)$ (when we couple the two graphs on the same probability space and define them on the same node set). Then via a transitive argument,  a uniform $q$-intersection graph $G_q(n, K_n,P_n)$ contains an Erd\H{o}s--R\'{e}nyi graph $G(n,y_n)$ as a spanning subgraph with probability $1-o(1)$ (when we couple the two graphs on the same probability space and define them on the same node set). Of course, we still need to show that (i) given the conditions of Lemma \ref{lem-cpgraph-rigrig}, all
conditions in Lemmas \ref{brig_urig} and \ref{er_brig} hold; and (ii)  $y_n$ defined in (\ref{pnpb01}) satisfies  (\ref{ERgraph-sn-defn-reduced}). Since the proofs are straightforward, we omit the details for simplicity.




\begin{lem} \label{brig_urig}
If $ K_n =
\Omega(n^{\epsilon})$ for a positive constant $\epsilon$, $ \frac{{K_n}^2}{P_n}  =
 o\left( \frac{1}{\ln n} \right)$,  and $ \frac{K_n}{P_n} = o\left( \frac{1}{n\ln n} \right)$, with
$x_n$ set by
\begin{align}
 x_n   = \textstyle{\frac{K_n}{P_n}
 \Big(1 - \sqrt{\frac{3\ln
n}{K_n }}\hspace{2pt}\Big)}, \label{pnKn}
 \end{align}
then 
 it holds that
\begin{align}
  G_q(n,K_n,P_n) & \succeq_{1-o(1)}H_q(n,x_n,P_n). \label{eq_brig_urig}
\end{align}

\end{lem}

\begin{lem} \label{er_brig}
If
 \begin{align}
\textstyle{ x_n  P_n } & = \Omega(n^{\epsilon}) \text{ for a positive constant $\epsilon$}, \label{er_brig-eq1} \\  \textstyle{ {x_n} } & = \textstyle{o\left( \frac{1}{n\ln n} \right)}, \label{er_brig-eq2} \\ \textstyle{{x_n}^2 P_n} &   =\textstyle{o\left( \frac{1}{\ln n} \right)}, \text{ and} \label{er_brig-eq3} \\ \textstyle{{x_n}^2 P_n} & = \textstyle{ \omega\big(\frac{(\ln n)^6}{n^2}\big)}, \label{er_brig-eq4}
 \end{align}
 then
there exits some $y_n$ satisfying
\begin{align}
y_n & = \textstyle{\frac{(P_n{x_n}^2)^q}{q!}} \cdot \big[1- o\left( \frac{1}{\ln n} \right)\big]
\label{pnpb01}
\end{align}
such that Erd\H{o}s--R\'{e}nyi graph $G(n,y_n)$
  obeys
\begin{align}
 H_q(n,x_n,P_n)& \succeq_{1-o(1)} G(n,y_n) . \label{GerGb}
\end{align}

\end{lem}

We can establish Lemmas \ref{brig_urig} and \ref{er_brig} in a way similar to that in \cite{ICASSP17-design}.
After establishing Lemmas \ref{brig_urig} and \ref{er_brig} to obtain Lemma \ref{lem-cpgraph-rigrig} and then using Lemma \ref{lem-cpgraph-rigrig} to get Lemma \ref{lem-cp_rig_er},
 we evaluate $z_n$ given by (\ref{ERgraph-sn-defn}) under the conditions of Theorem \ref{thm:OneLaw+NodeIsolation}. First, as explained in Section \ref{sec-Confining}, to prove Theorem \ref{thm:OneLaw+NodeIsolation}, we can introduce the extra condition $|\alpha_n |= o(\ln n)$. Then under the conditions of Theorem \ref{thm:OneLaw+NodeIsolation} with the extra condition $| \alpha_n | = o(\ln n)$, we can show that all conditions of Lemma \ref{lem-cpgraph-rigrig}  hold, and $z_n$ given by (\ref{ERgraph-sn-defn}) satisfies
\begin{align}
\textstyle{z_n   = \frac{\ln  n + {(k-1)} \ln \ln n + {\alpha_n}-o(1)}{n}.} \label{ER-sn-kcon}
\end{align}
For $z_n$ satisfying (\ref{ER-sn-kcon}), we obtain from Lemma \ref{lem:ER:kcon} below that probability of $G(n,z_n)$ being $k$-connected
 can be written as $e^{-
\frac{e^{- \iffalse \lim\limits_ \fi \lim_{n \to \infty}{\alpha_n}}}{(k-1)!}} \cdot [1\pm o(1)]$, where we use $\lim_{n \to \infty}[{\alpha_n}-o(1)] = \lim_{n \to \infty}{\alpha_n}$. This result and (\ref{Gq-kcon-lower-bound-tonsb}) further induce that $\mathbb{G}_{n,q}$ under the conditions of Theorem \ref{thm:OneLaw+NodeIsolation} with $| \alpha_n | = o(\ln n)$ is $k$-connected with probability at least $e^{-
\frac{e^{- \iffalse \lim\limits_ \fi \lim_{n \to \infty}{\alpha_n}}}{(k-1)!}} \times [1- o(1)]$. This proves the lower bound in Section \ref{sec-Decomposing}.

\begin{lem}[\textbf{$k$-Connectivity in an Erd\H{o}s--R\'{e}nyi graph} by {\cite[Theorem 1]{erdoskcon}}\hspace{0pt}] \label{lem:ER:kcon}
For an Erd\H{o}s--R\'enyi graph $G(n,z_n)$, if there is a sequence $\alpha_n$ with $\lim_{n \to \infty}{\alpha_n} \in [-\infty, \infty]$
such that $z_n  = \frac{\ln  n    + {(k-1)} \ln \ln n  +
 {\alpha_n}}{n}$,
 then it holds that
 \begin{align}
 \lim_{n \to \infty}   \mathbb{P}[ G(n,z_n)\text{ is $k$-connected.} ] &  = e^{- \frac{e^{-\lim_{n \to \infty}{\alpha_n}}}{(k-1)!}} . \nonumber
 \end{align}
 \end{lem}

\section{Proving the Upper Bound of   Section \ref{sec-Decomposing}} \label{sec-prf-upper-bounds-details}

The idea to prove the upper bound $e^{-
\frac{e^{- \iffalse \lim\limits_ \fi \lim_{n \to \infty}{\alpha_n}}}{(k-1)!}} \times [1 + o(1)]$ for $\mathbb{P}[\hspace{2pt}\mathbb{G}_{n,q}\text{ is
  $k$-connected.}\hspace{2pt}]$ has been explained in Section \ref{sec-prf-upper-bounds}. As explained, we derive the asymptotically exact probability for the property of minimum   degree being at least $k$ in the studied graph $\mathbb{G}_{n,q}$. The result is presented as Lemma \ref{lem-mnd} below,  where $t (K_n, P_n,q, {p_n})$ (i.e., $t_{n,q}$ in short) is the edge probability of $\mathbb{G}_{n,q}$. Note that the conditions of Lemma \ref{lem-mnd} all hold under the conditions of Theorem \ref{thm:OneLaw+NodeIsolation}.

\begin{lem}[\textbf{Property of minimum degree being at least $k$ in graph $\mathbb{G}_{n,q}\iffalse_{on}^{(q)}\fi$}]\label{lem-mnd}

For a graph $\mathbb{G}_{n,q}(n, K_n,P_n, {p_n})$, if there exists a sequence $\alpha_n$ with $\lim_{n \to \infty} \alpha_n \in [-\infty, +\infty]$ such that
\begin{align}
t (K_n, P_n,q, {p_n})  & = \frac{\ln  n + (k-1) \ln \ln n   +
 {\alpha_n}}{n}  , \label{lem-mnd-t-edgeprob}
\end{align}
then it holds under $ K_n =
\Omega(n^{\epsilon})$ for a positive constant $\epsilon$, $ \frac{{K_n}^2}{P_n}  =
 o\left( \frac{1}{\ln n} \right)$,  and $ \frac{K_n}{P_n} = o\left( \frac{1}{n\ln n} \right)$ that
\begin{align}
& \hspace{-27pt}  \lim_{n \rightarrow \infty } \bP{\text{$\mathbb{G}_{n,q}\iffalse_{on}^{(q)}\fi$  has a minimum degree at least $k$.}}
 \nonumber
\\ &  \hspace{-27pt}  =  e^{- \frac{e^{-\lim_{n \to \infty} \alpha_{_n}}}{(k-1)!}}
\end{align}
\vspace{-5pt}
\begin{subnumcases}
{\hspace{-27pt} =\hspace{-2pt}} \hspace{-3pt}e^{- \frac{e^{-\alpha ^*}}{(k-1)!}},
 &\hspace{-11.5pt}\text{ if $\lim\limits_{n \to \infty}{\alpha_n}
=\alpha ^* \in (-\infty, \infty)$,}\\  \hspace{-3pt}1, &\hspace{-11.5pt}\text{ if $\lim\limits_{n \to \infty}{\alpha_n}
=\infty$,} \label{thm-mndx-eq-1}  \\ \hspace{-3pt} 0, &\hspace{-11.5pt}\text{ if $\lim\limits_{n \to \infty}{\alpha_n}
=-\infty$}.\label{thm-mndx-eq-0} \label{thm-mndx-eq-e}
\end{subnumcases}

\end{lem}



 We establish Lemma \ref{lem-mnd} for minimum degree in graph $\mathbb{G}_{n,q}$ by analyzing the asymptotically exact distribution for the number of nodes with a fixed degree, for which we present Lemma \ref{thm:exact_qcomposite2} below.

 The details of using Lemma \ref{thm:exact_qcomposite2} to prove Lemma \ref{lem-mnd} are given in \cite{ToN17-topological}. We   show that to prove Lemma \ref{thm:exact_qcomposite2}, the deviation $\alpha_n$ in the lemma statement can  be confined as $\pm  o(\ln n)$. More specifically, if Lemma \ref{thm:exact_qcomposite2} holds under the extra condition $|\alpha_n |= o(\ln n)$, then Lemma \ref{thm:exact_qcomposite2} also holds regardless of the extra condition. For constant $k$ and   $|\alpha_n |= o(\ln n)$, clearly $t (K_n, P_n,q, {p_n})$ in (\ref{lem-mnd-t-edgeprob}) satisfies (\ref{peq1}).

\begin{lem}[\textbf{Possion distribution for number of nodes with a fixed degree in graph $\mathbb{G}_{n,q}\iffalse_{on}^{(q)}\fi$}]  \label{thm:exact_qcomposite2}
For graph $\mathbb{G}_{n,q}$ with $ K_n =
\Omega(n^{\epsilon})$ for a positive constant $\epsilon$, $ \frac{{K_n}^2}{P_n}  =
 o\left( \frac{1}{\ln n} \right)$,  and $ \frac{K_n}{P_n} = o\left( \frac{1}{n\ln n} \right)$, if
\begin{align}
t (K_n, P_n,q, {p_n}) & = \frac{\ln  n \pm o(\ln n)}{n},  \label{peq1}
\end{align}
then for a non-negative constant integer $h$, the number of nodes in
$\mathbb{G}_{n,q}$ with degree $h$ is in distribution asymptotically equivalent to a
Poisson random variable with mean $\lambda_{n,h} : = n (h!)^{-1}(n t_{n,q})^h e^{-n
t_{n,q}}$, where $t_{n,q}$ is short for $t (K_n, P_n,q, {p_n}) $; i.e., as $n \to \infty$,
\begin{align}
& \hspace{-6pt}\bP{\hspace{-3pt}\begin{array}{l}\text{The number of nodes in
$\mathbb{G}_{n,q}\iffalse_{on}\fi$}\\\text{with degree $h$ equals $\ell$.}\end{array}\hspace{-3pt}}\hspace{-2pt}\Big/\hspace{-2pt}\Big[(\ell !)^{-1}{\lambda_{n,h}} ^{\ell}e^{-\lambda_{n,h}}\Big]\hspace{-2pt} \to\hspace{-2pt} 1, \nonumber \\ & \text{~~~~~~~~~~~~~~~~~~~~~~~~~~~~~~~~~~~~~~~~~~~for $\ell = 0,1, \ldots$}  \label{eq-Poisson-lemma}
\end{align}
\end{lem}

Lemma \ref{thm:exact_qcomposite2} for graph
$\mathbb{G}_{n,q}$ shows that the number of nodes
with a fixed degree follows a Poisson distribution
asymptotically. Lemma \ref{thm:exact_qcomposite2} is established in \cite{ToN17-topological}.

\section{Related Work} \label{related}

Graph ${G}_q(n, K_n, P_n)$ models
the topology of a secure sensor network with the $q$-composite key predistribution scheme under full visibility, where full visibility
means that any pair of nodes have active channels in between so the only requirement for secure communication
is the sharing of at least $q$ keys.
For graph $G_q(n, K_n, P_n)$, Bloznelis and {\L}uczak \cite{Perfectmatchings} (resp., Bloznelis and Rybarczyk
\cite{Bloznelis201494}) have recently derived the asymptotically exact
probability for $k$-connectivity (resp., connectivity). The result of \cite{Bloznelis201494} is also obtained by Zhao \emph{et al.} \cite{ANALCO} under more general conditions.

 Zhao \emph{et al.} \cite{ANALCO} have recently derived a zero--one law for $k$-connectivity in $G_q(n, K_n, P_n)$. With $s(K_n, P_n, q)$ being the edge probability of $G_q(n, K_n, P_n)$, they show that under $ P_n =
\Omega(n)$, with $\alpha_n$ defined through
$ s(K_n, P_n, q)  = \frac{\ln  n   + (k-1)\ln\ln n+
 {\alpha_n}}{n}$, then $G_q(n, K_n, P_n)$ is $k$-connected with probability $e^{- \frac{e^{-\alpha ^*}}{(k-1)!}}$ if $\lim\limits_{n \to \infty} \alpha_n = \alpha ^*$, is not $k$-connected  with high probability if $\lim_{n \to \infty}{\alpha_n} = -\infty$, and is $k$-connected with high probability if $\lim_{n \to \infty}{\alpha_n}
 = \infty$.  Other properties of $G_q(n, K_n, P_n)$ are also considered in the literature. For example, Bloznelis \emph{et al.}
\cite{Rybarczyk} demonstrate that a connected component with at at
least a constant fraction of $n$ emerges with high probability when
the edge probability $s(K_n, P_n, q)$ exceeds $1/n$. Nikoletseas \emph{et al. } \cite{NikoletseasHM} investigate Hamilton cycles in $G_q(n, K_n, P_n)$, where a Hamilton cycle in a graph is a closed loop that visits each node once. When $q=1$, graph ${G}_1(n, K_n, P_n)$ models
the topology of a secure sensor network with the Eschenauer--Gligor key predistribution scheme under full visibility. For ${G}_1(n, K_n, P_n)$, its connectivity
  has been investigated extensively \cite{ryb3,r1,yagan,DiPietroMeiManciniPanconesiRadhakrishnan2004}. In particular, Di Pietro \emph{et al.} \cite{DiPietroMeiManciniPanconesiRadhakrishnan2004} show that under $K_n = 2$ and \vspace{2pt} $P_n = \frac{n}{\ln n}$, graph
  $G_1(n, K_n, P_n)$ is connected with high probability; Di Pietro \emph{et al.} \cite{DiPietroTissec} establish
  that under $P_n \geq n$ and $\frac{{K_n}^2}{P_n} \sim \frac{\ln n}{n}$,
  $G_1(n, K_n, P_n)$ is connected with high probability;
 and Ya\u{g}an and Makowski \cite{yagan} prove that
  under $ P_n =
\Omega(n)$, with $\alpha_n$ defined by
$\frac{{K_n}^2}{P_n} = \frac{\ln  n   +
 {\alpha_n}}{n}$, then $G_1(n, K_n, P_n)$ is  disconnected with high probability if $\lim_{n \to \infty}{\alpha_n}
 = -\infty$ and connected with high probability if $\lim_{n \to \infty}{\alpha_n}
 = \infty$. For $k$-connectivity in  $G_1(n,K_n,P_n)$,  Rybarczyk \cite{zz}
implicitly shows a zero--one law, and we \cite{ZhaoTAC} derive the asymptotically exact probability.

  Erd\H{o}s and R\'{e}nyi \cite{citeulike:4012374} 
 introduce the random graph model $G(n,p_n)$ defined on
a node set with size $n$ such that an edge between any two nodes
exists with probability $p_n$ \emph{independently} of all other
edges. Graph $G(n,p_n)$ models the topology induces by a sensor
network under the on/off channel model of this paper. Erd\H{o}s and R\'{e}nyi
\cite{citeulike:4012374} (resp., \cite{erdoskcon})    derive a zero--one law for connectivity (resp., $k$-connectivity) in
graph $G(n,p_n)$; specifically, the result of \cite{erdoskcon} is that with $\alpha_n$ defined through
$ p_n  = \frac{\ln  n    + (k-1)\ln\ln n+
 {\alpha_n}}{n}$, then $G(n,p_n)$ is not $k$-connected  with high probability if $\lim_{n \to \infty}{\alpha_n}
 = -\infty$ and $k$-connected with high probability if $\lim_{n \to \infty}{\alpha_n}
 = \infty$.

As detailed in Section
\ref{sec:SystemModel}, the graph model $\mathbb{G}_{n,q}(n, K_n,P_n, {p_n}) = G_q(n, K_n, P_n) \cap G(n,p_n)$ studied in this paper represents the topology of a secure sensor network employing the $q$-composite key predistribution scheme
\cite{virgil} under the on/off channel model.
For graph $\mathbb{G}_{n,q}(n, K_n,P_n, {p_n}) $, Zhao \emph{et al.} \cite{ToN17-topological,zhao2014topological} have recently studied its node degree distribution, but not connectivity. When $q=1$, graph $\mathbb{G}_{n,q}(n, K_n,P_n, {p_n}) $ reduces to $\mathbb{G}_1(n, K_n,P_n, {p_n}) $, which models the topology of a secure sensor network employing the Eschenauer--Gligor key predistribution scheme under the on/off channel model. For graph $\mathbb{G}_1(n, K_n,P_n, {p_n}) $, Ya\u{g}an \cite{yagan_onoff} presents a zero--one law for
connectivity. With $ s(K_n, P_n, 1)$ being the edge probability of $G_1(n, K_n, P_n)$ and hence $ s(K_n, P_n, 1) \cdot p_n$ being the edge probability of $\mathbb{G}_1(n, K_n,P_n, {p_n}) = G_1(n, K_n, P_n) \cap G(n,p_n)$, Ya\u{g}an \cite{yagan_onoff} shows that
under $P_n = \Omega(n)$, $\frac{{K_n}^2}{P_n} = o(1)$, the existence of
$\lim_{n\to \infty}({p_n\iffalse_{on}\fi}\ln n)$ and
$s(K_n, P_n, 1) \cdot p_n \sim \frac{c\ln n}{n} $
for a positive constant $c$,
then graph
$G_1(n, K_n, P_n)$ is disconnected with high probability if $c<1$ and connected with high probability if $c>1$. Zhao \emph{et al.} \cite{ZhaoYaganGligor} extend Ya\u{g}an's result \cite{yagan_onoff} on  connectivity to $k$-connectivity with a more fine-grained scaling. Graph $\mathbb{G}_{n,q}(n, K_n,P_n, {p_n}) $ with general $q$ has also recently been studied in the literature:  \cite{GlobalSIP15-parameter} presents the exact probability result of connectivity, while \cite{ICASSP17-social} derives a
zero--one law for $k$-connectivity. This paper provides the exact probability result of $k$-connectivity in $\mathbb{G}_{n,q}(n, K_n,P_n, {p_n}) $. Obtaining the exact probability result rather than just a zero--one law provides more precise guidelines for the design of secure sensor networks employing   $q$-composite key predistribution  with   on/off channels.

The analysis of secure sensor networks has also been considered under
physical link constraints different with the on/off
channel model, where one example is the popular
disk model \cite{4428764,zhao2015resilience,Pishro}. In the disk model,
  nodes are distributed over a bounded region of a Euclidean
plane, and two nodes have to be within a certain distance for
communication. Although several studies \cite{ZhaoAllerton,Krzywdzi,Pishro,DiPietroMeiManciniPanconesiRadhakrishnan2004,4151628,ISIT_RKGRGG} have  investigated connectivity in secure sensor networks
under the disk model, a zero--one law (that is similar to
our result under the on/off channel model) for $k$-connectivity
in secure sensor networks  employing the
$q$-composite key predistribution scheme under the disk model remains an open question. However, a zero--one law
similar to our result here is expected to hold in view of the similarity in ($k$-)connectivity between the random graphs
 induces by the disk model \cite{zhao2015resilience} and the on/off channel model \cite{yagan_onoff}.

\section{Conclusion} \label{sec:Conclusion}

In this paper, we present the asymptotically exact probability and a zero--one law for $k$-connectivity
 in a secure wireless sensor network operating under the
$q$-composite key predistribution scheme with on/off channels. The
network is modeled by composing a uniform $q$-intersection graph with
an Erd\H{o}s-R\'enyi
graph, where the uniform $q$-intersection graph characterizes the $q$-composite key predistribution scheme
and the Erd\H{o}s-R\'enyi
graph captures the on/off channel model. Experimental results are
shown to be in agreement with our theoretical findings.

\end{document}